# Size-Limited Room Temperature Single-Photon Emission from Sidewall-Treated Fractional Dimension InGaN Quantum Dots: Determined by Density-of-States-Corrected Ultrafast Carrier Dynamics and Improved Signal-to-Noise Ratio

*Pratim K. Saha**

Institute of Physics and Astronomy

Technical University of Berlin

Hardenbergstraße 36, 10623 Berlin, Germany

*Email- p.k.sahaelectronics@gmail.com

Room-temperature single-photon emission (SPE) resulting from a biexciton-exciton cascaded decay is demonstrated for the first time from chemically and photoelectrochemically etched site-controlled $In_{0.14}Ga_{0.86}N$ quantum dots (QDs) embedded in vertical GaN nanowires. Diameter-dependent biexciton-exciton dynamics are analysed to determine the eligibility of QD as a single-photon emitter. The signal-to-noise ratio degrades with increasing QD diameter. Background noise photons pose a bottleneck to achieving SPE. This is also explained from a carrier dynamics perspective. Surface recombination contributes to inhomogeneous broadening at QD diameters > 35 nm. Below 35 nm, density-of-states-corrected Auger gradually becomes the principal biexciton-decay route with further reduction in QD diameter, thereby quenching the possibility of thermal broadening and setting a threshold for SPE. Below 9 nm, the Auger recombination rate becomes manyfold of other decay rates, causing multi-photon suppression via single Auger decay to form an exciton. Surface recombination probability of this exciton is minimized while biexciton state filling probability is maximized by reducing sidewall surface states through wet-treatment. These improve biexciton state preparation and enhance the single-photon purity of the exciton towards the exciton Bohr radius (3 nm) regime. Far away from this regime, higher-order autocorrelations to characterize quantum emission involving multi-photon events are discussed. This study establishes a generalized physical framework for predetermining SPE probability as a function of QD's surface and geometry down to the exciton Bohr radius regime, with practical implementations. This work shows the pathway to design and develop next-generation semiconductor QDs for high-purity room-temperature SPE.

**Keywords**- Single-photon emission, density-of-states-corrected Auger recombination, surface recombination, fractional dimension, InGaN quantum dot, wet-chemical etching, higher-order autocorrelation

## I. INTRODUCTION

Semiconductor quantum dots (QDs) are used as single-photon emitters in different quantum technology platforms. Quantum information is encoded in polarization, time or phase bin of the emitted photons, and it is further coupled, transmitted, processed and decoded for practical applications such as quantum computation, quantum cryptography, quantum communication and quantum metrology[1-17]. Rydberg atoms, diamond colour centres, and single molecules are also well-known contemporary systems to generate single photons on demand[18-21]. Alternatively, heralded single photons can be generated via spontaneous parametric down-conversion (SPDC)[22]. Among these, solid-state QD single-photon sources are the most promising and practical choice due to their GHz clock rate, easy integration with nanophotonic waveguide cavities, scalability, high-temperature robustness, and compatibility with optical fibre for long-distance quantum communication with low transmission loss[14, 22-27]. Nitride QDs are almost free of material defects[28, 29] and facilitate sub-nm emission linewidth along with sub-ns radiative decay time even at room temperature[30-36]. In addition, site-controlled QDs fabricated using a top-down approach are uniform in size and shape compared to epitaxially grown QDs and hence more beneficial for generating indistinguishable photon pairs. Though there are many reports on SPE via biexciton-exciton cascaded decay from site-controlled and self-assembled grown QDs[37-39], there are hardly any reports on the same from chemically and photoelectrochemically etched site-controlled QDs. To the best of my knowledge, there are no reports on the quantum optical properties of semiconductor QDs as a function of QD size and surface properties as well. Moreover, a study of size-dependent Auger recombination rates in fractional dimension InGaN QDs is also missing. In this work, the quantum optical properties and ultrafast carrier dynamics of chemically and photo-electrochemically (PEC) treated and laterally-etched $In_{0.14}Ga_{0.86}N$ QDs with a height of 3 nm and varying diameters are analysed in detail. The diameters of QDs are measured from their cross-sectional scanning electron microscopy (cross-SEM) images for chemically etched QDs and calculated using micro-photoluminescence (µPL) results for PEC etched QDs. The limiting conditions for single-photon emission (SPE) are investigated separately in the context of signal-to-noise ratio (SNR) extracted from room-temperature µPL spectra and carrier recombination dynamics.

## II. MATERIALS AND METHODS

### A. Sample Preparation

The InGaN/GaN single-quantum well (3 nm)/barrier (12 nm) heterostructure is grown along the c-axis on a sapphire substrate by the metal organic chemical vapour deposition (MOCVD) technique. A 30 nm-diameter circular pattern is transferred via electron-beam lithography (EBL). Subsequently, the sample is dry-etched using inductively coupled plasma reactive ion etching (ICPRIE) at 500 V DC, and 75 V RF bias for 100 s, maintaining the $BCl_3/Cl_2/Ar$ gas flow rates at 5/30/20 sccm. Four such dry-etched samples (S1) are initially prepared. Two of these samples are PEC etched for 40 minutes (S2) and 2 hours (S5) in $H_2SO_4$ (1.2 %) solution under an optical bath of 450 mW broadband light (250-800 nm) source in the presence of an external bias. The recipe for this etching is provided in Table S1. The other two samples are chemically etched using boiling TMAH at 90 $^0$C for 28 (S3) and 35 minutes (S4). The detailed frameworks of the PEC and TMAH etching are discussed in my previous work[40].

## III. RESULTS AND DISCUSSION

### A. Scanning Electron Microscopy

InGaN/GaN single quantum well (3 nm)/barrier (12 nm) heterostructure grown on a c-plane sapphire platform is used in this work. A schematic of the heterostructure is shown in Figure 1. QDs are fabricated from a quantum well sample (S1) using a combination of dry and wet etching techniques. After dry etching, QD samples S2 and S5 are prepared by means of PEC etching, while QD samples S3 and S4 are prepared via TMAH etching. Cross-sectional SEM images of the samples are shown in Figure 2. QD diameters of 36 and 30 nm are measured for samples S3 and S4, respectively. For PEC-etched samples S2 and S5, the QD diameters can not be measured directly from their SEM images. Instead, they are calculated from the µPL peak shifts of the final QDs relative to their dry-etched versions. This calculation scheme is discussed in detail in my previous work[40].

### B. Micro-photoluminescence

The µPL spectra of the samples are shown in Figure 3a. The µPL responses are measured under 405 nm laser excitation at room temperature. InGaN excitonic emission peaks (X) resulting from biexciton-exciton cascaded decay, along with background defect peaks (D), are present in the spectra for all QD samples. The biexciton peak (XX) corresponding to biexciton-to-exciton radiative decay is present for samples S2, S3 and S4, but not in S5. The possible reason behind the absence of the biexciton peak for sample S5 is explained in section 3.5 below. For sample S4, the exciton and biexciton peaks overlap, and they are individually fitted using the Lorentz function (shown in the inset of panel S4 in Figure 3a). The separations between the exciton and biexciton peaks are 2.9, 3 and 0.7 nm for samples S2, S3 and S4, respectively. The linewidths of the exciton and biexciton peaks are mentioned in Figure 3a for all QD samples, wherever applicable. An increase in the exciton as well as the biexciton linewidth is observed with QD diameter. This linewidth broadening can be attributed to additional phonon sidebands arising from non-radiative thermal recombination, as discussed in detail in section 3.5 below. Exciton peaks vary between 430-435 nm for samples S2, S3 and S4 and show a strong blue-shift of 22 nm for sample S5. This is the signature of a strong quantum confinement in S5. QD diameters for samples S2 and S5, calculated from their respective µPL peak-shifts, are 36 and 8 nm, respectively. Defect emission peaks are observed at 550 nm across all samples. The defect peak for sample S5 is merely visible when plotted with its exciton peak due to the large SNR. Hence, it is separately shown in the inset of panel S5 (Figure 3a).

### C. Autocorrelation Measurements

The second-order autocorrelation of excitonic emission is experimentally measured for each QD sample using a Hanbury-Brown-Twiss (HBT) setup. Laser excitation and collection of QD emission are performed in a confocal arrangement. A schematic of the experimental setup is shown in my previous work[36]. A spectral filter was used during the experiment to block residual 405 nm excitation light in the detection path, but it neither blocked the defect emission nor suppressed the biexciton photons. Experimentally measured autocorrelations of the samples are shown in Figure 3b. SPE is observed from samples S4 and S5, whereas antibunched multi-photon emission is recorded from samples S2 and S3. Signatures of blinking, along with higher background noise, are evident for samples S4 and S5. The smaller InGaN volume in these samples, relative to S2 and S3, results in lower signal counts, thereby making the detector's dark noise visible. The green and blue eye-guiding envelopes are used for samples S4 and S5, respectively, to clearly indicate the zero-delay and side peaks in Figure 3b.

### D. Calculation of $g^{(2)}(0)$ from µPL Signal-to-Noise Ratio: Limiting Condition for SPE

Zero-delay autocorrelation ($g^{(2)}(0)$) of the excitonic emission is calculated using two different approaches. The first method calculates $g^{(2)}(0)$ from the SNR (r) of µPL spectra. The $g^{(2)}(0)$ can be expressed as[41]-

$$g^{(2)}(0) = \frac{1+2r}{(r+1)^2} \qquad (1)$$

Where, $\qquad r = \frac{S}{B} \qquad (2)$

S is the integrated exciton peak, whereas B is the sum of the integrated biexciton peak and the integrated defect peak.

Background correction factor (ρ) is defined as[42],

$$\rho = \frac{S}{S+B} \quad (3)$$

Hence, $g^{(2)}(0)$ can be written in terms of ρ as,

$$g^{(2)}(0) = 1 - \rho^2 \quad (4)$$

For a single-photon source, $g^{(2)}(0)$ should be less than 0.5, the single-photon limit (SPL). This, in turn, puts a restriction on the value of ρ ($\rho_{SPL} = \sqrt{0.5}$) and r ($r_{SPL} = 2.414$) below which the SPE property ceases. The $g^{(2)}(0)$ is plotted as a function of r and ρ in Figure 3c. The $g^{(2)}(0)$ values calculated from μPL spectra are indicated by discrete points in the figure. It is observed that samples S4 and S5 exhibit single-photon behaviour. On the contrary, S2 and S3 do not show single-photon properties, which can be ascribed to comparatively poor SNR for these samples.

E.  **Limiting Conditions for SPE from Carrier-Carrier Interaction Energy and Dynamics**

Zero-delay second-order autocorrelation is further calculated using the second method. Since the emission occurs through a cascade decay process of biexciton and exciton, $g^{(2)}(0)$ can be presented as the ratio of the quantum efficiency (η) of the biexciton-to-exciton to that of the exciton-to-ground state recombination. Quantum efficiency, in each case, is defined as the ratio of the radiative recombination rate to the total recombination rate. This approximation has been used previously[43] and holds true for an InGaN QD system used in this work. Here, I consider exciton decay to the ground state through radiative and non-radiative channels, including tunnelling and thermally induced surface recombination. For biexciton, carriers have an additional Auger escape route. The schematics of carrier decay processes are shown in Figures 4a and 4b for QDs with a diameter larger than 35 nm and smaller than 9 nm, respectively. The Shockley-Read-Hall (SRH) recombination is a slower process, especially at higher carrier concentrations, as previously reported for InGaN quantum wells and QDs[44, 45]. Apart from this, SRH, being a trap-assisted volume recombination process, has a much lower probability in a small volume and in an almost defect-free system like a site-controlled QD. Hence, it has been neglected in my calculation. The $g^{(2)}(0)$ can be written as follows[43],

$$g^{(2)}(0) = \frac{\eta_{XX}}{\eta_X} \tag{5}$$

Where,
$$\eta_{XX} = \frac{\gamma_{r,XX}}{\gamma_{r,XX} + \gamma_{nr,XX}} \tag{6}$$

And,
$$\eta_X = \frac{\gamma_{r,X}}{\gamma_{r,X} + \gamma_{nr,X}} \tag{7}$$

$\gamma_{r,X}, \gamma_{nr,X}, \gamma_{r,XX}$ and $\gamma_{nr,XX}$ are radiative and non-radiative recombination rates of exciton and those of biexciton, respectively, and the inverse of them are their respective recombination time constants. Hence, the term $g^{(2)}(0)$ can be rewritten as,

$$g^{(2)}(0) = \frac{1 + \frac{\tau_{r,X}}{\tau_{nr,X}}}{1 + \frac{\tau_{r,XX}}{\tau_{nr,XX}}} \tag{8}$$

$\tau_{r,X}, \tau_{r,XX}, \tau_{nr,X}$ and $\tau_{nr,XX}$ are the radiative time constants of exciton, biexciton and non-radiative time constants of exciton, biexciton, respectively. I shall first discuss the calculation for non-radiative processes, followed by the same for the radiative rate. The non-radiative time constants can be expressed as,

$$\frac{1}{\tau_{nr,X}} = \frac{1}{\tau_{tunnel,X}} + \frac{1}{\tau_{thermal,X}} \tag{9}$$

And,
$$\frac{1}{\tau_{nr,XX}} = \frac{1}{\tau_{tunnel,XX}} + \frac{1}{\tau_{thermal,XX}} + \frac{1}{\tau_{Auger,XX}} \tag{10}$$

For a single-photon emitter,
$$g^{(2)}(0) < 0.5 \tag{11}$$

Hence, substitution of $g^{(2)}(0)$ from equation (8) into the inequality (11) yields,

$$\frac{\tau_{r,XX}}{\tau_{nr,XX}} > \frac{2\tau_{r,X}}{\tau_{nr,X}} + 1 \tag{12}$$

Tunnelling and thermal processes take place near the QD sidewall, where carriers experience a radial potential barrier. This barrier forms automatically due to strain relaxation in the QD near its sidewall. The radial strain-relaxation profile is calculated using COMSOL for InGaN QDs with diameters ranging from 4 to 36 nm. Corresponding potential profiles are computed by solving the 2D Schrödinger-Poisson equation, considering cylindrical symmetry in Silvaco. The thermal[43], tunnelling[43] and Auger[46] recombination rates of the exciton (biexciton) are defined as follows,

$$\tau_{\text{thermal, X(XX)}} = \frac{D}{K_1 e^{\frac{-K_2 \Phi_{B,X(XX)}}{KT_1}}} \quad (13)$$

$$\tau_{\text{tunnel, X(XX)}} = \frac{D}{K_3 e^{-K_4 \int_0^{D/2} \sqrt{\Phi(r)_{X(XX)}} dr}} \quad (14)$$

$$\tau_{\text{Auger, XX}} = \frac{D^7 E_g^{\frac{7}{2}}}{K_5 (E_g^m - E_g)} \quad (15)$$

Where D is QD diameter, $\Phi(r)$ is the radial potential profile, $\Phi_B$ is the maximum potential barrier, $E_g^m$ is the bandgap of barrier material, $E_g$ is the InGaN QD transition energy, $T_1$ is the ambient working temperature, $K$ is the Boltzmann constant and $K_{i\,(i=1-4)}$ is the material constant.

The biexciton potential is calculated using the equation below,

$$\Phi_{B,\,XX} = \Phi_{B,\,X} - E_{B,\,XX} \quad (16)$$

Biexciton binding energy is obtained from the following equation[47],

$$E_{B,\,XX} = 2E_{B,\,X} - E_{e-e} - E_{h-h} \quad (17)$$

Alternatively, it can also be written as follows-

$$E_{B,\,XX} = E_X - E_{XX} \quad (18)$$

Where $E_{B,\,X}$, $E_{e-e}$ and $E_{h-h}$ are the exciton binding energy, electron-electron and hole-hole electrostatic repulsive energy, respectively. $E_X$ and $E_{XX}$ denote, respectively, the exciton and biexciton energy states relative to the ground state.

The Auger recombination time constant calculated as a function of QD diameter using equation (15) yields very small and unrealistic time constants for smaller diameters. The rate of change of the time constant with diameter is also considerably high. Consideration of the decrease in density of states (DOS) with QD diameter compensates for the Auger term and yields a realistic result. Due to reduced DOS in lower dimensions, the probability of an Auger recombination event decreases, thereby increasing the time constant for any particular QD diameter after correction. The corrected expression of the Auger recombination time constant is given by,

$$\tau_{Auger,\ XX(corrected)} = \frac{D^7 E_g^{\frac{7}{2}}}{K_5 (E_g^m - E_g) DOS_{frac}} \quad (19)$$

$$DOS_{frac} = 0.52(1-T)\pi^{1-T} 10^{18} 2^{2T-2} D^{2T} \quad (20)$$

$$T = \frac{\Delta E_c - \Delta E_{2D}}{\Delta E_{0D} - \Delta E_{2D}} \quad (21)$$

Where, $\Delta E_c$, $\Delta E_{0D}$ and $\Delta E_{2D}$ are the differences between the bottom edge of the conduction band for bulk $In_{0.14}Ga_{0.86}N$ and the first electron bound state energy for the fractional, 0D and 2D cases, respectively.

This model is further verified by calculating the Auger recombination coefficients using the diameter-dependent corrected Auger rates reported in the present work. An average Auger coefficient of $2\times10^{-30}$ $cm^6 s^{-1}$, as obtained, is consistent with previously reported values for the same material system[48-51].

To evaluate the exciton's radiative lifetime, I considered the previously reported diameter-dependent radiative lifetime of the polar InGaN QD with a 2.5 nm height, and then calculated the same for a QD with 3 nm height, using the height-dependent radiative lifetime plot[52]. Previous studies reported a dependence of the exciton-to-biexciton lifetime ratio on QD aspect ratio[53, 54]. For a QD with lateral (d) as well as vertical (V) dimension in the range of exciton Bohr radius ($a_0$ = 3 nm), this ratio is 2[53, 54]. If the lateral dimension is increased to 3d while keeping the other dimension constant at V, this ratio becomes 1.4[53, 54]. For lateral dimension>> $a_0$, this ratio was reported to be 1[55, 56]. Biexciton lifetimes are calculated from these ratios.

All of the recombination time constants so obtained are plotted together in Figure 5a with QD diameter. Biexciton radiative time constants based on previously reported exciton-to-biexciton lifetime ratios are shown as discrete points in the figure. These points are connected by a dashed line. The biexciton lifetimes for any other lateral dimensions of QDs are extracted from this line. Thermal recombination and tunnelling rates show an increase with a reduction in QD diameter. A decrease in the in-plane potential barrier height with QD diameter is responsible for this increment. The higher Auger recombination rate at smaller diameters is primarily due to very high carrier density, reduced surface-state-related carrier capture, and a higher probability of biexciton state filling. On the other hand, larger electron-hole spatial overlap in smaller QDs leads to an increased radiative decay rate. The Auger recombination time constant is the minimum for any QD diameter below 35 nm and effectively determines $g^{(2)}(0)$ in equation (8). Rate constants of thermal and tunnelling processes are almost similar for exciton and biexciton, as evident from Figure 5a. Hence, equations (9) and (10) can be related as-

$$\frac{1}{\tau_{nr,\ XX}} = \frac{1}{\tau_{nr,\ X}} + \frac{1}{\tau_{Auger,\ XX}} \tag{22}$$

Substituting $\tau_{nr,\ XX}$ from equation (22) into the inequality (12)-

$$\frac{1}{\tau_{nr,\ X}}\left(\frac{2\tau_{r,\ X}}{\tau_{r,\ XX}}-1\right) < \frac{1}{\tau_{Auger,\ XX}} - \frac{1}{\tau_{r,\ XX}} \tag{23}$$

This is the general condition that must be fulfilled to achieve SPE from an InGaN quantum dot. Now, for QDs with all its dimensions equivalent to $a_0$, the above inequality takes the following form under the consideration of $\frac{\tau_{r,\ X}}{\tau_{r,\ XX}} = 2$

$$\tau_{Auger,\ XX} < \frac{\tau_X}{2 + \frac{\tau_X}{\tau_{nr,\ X}}} \tag{24}$$

Where the net exciton recombination time constant $\tau_X = \frac{\tau_{r,X} \tau_{nr,X}}{\tau_{r,X} + \tau_{nr,X}}$ (25)

Since, $\tau_{nr,X} \ll \tau_{r,X}$ $\tau_X \approx \tau_{nr,X}$

Hence, $\tau_{Auger,XX} < \frac{\tau_{nr,X}}{3}$ (26)

When QD diameter $d \gg a_0$, $\frac{\tau_{r,X}}{\tau_{r,XX}} = 1$

The inequality (23) above can be written as follows- $\tau_{Auger,XX} < \tau_{nr,X}$ (27)

Inequalities (26) and (27) above set the conditions of SPE from $In_{0.14}Ga_{0.86}N$ QDs under strong and weak lateral confinement cases, respectively. It is evident from Figure 5a that the calculated Auger recombination rate below 9 nm diameter QDs satisfies inequality (26). Hence, biexciton recombination is accelerated and governed by the Auger process, which leads to exciton formation. This decay of biexciton through a non-radiative channel minimizes the requirement of spectral filtering of biexciton photons. The exciton, so formed, subsequently decays through different possible recombination channels discussed above. Though thermal and tunnelling processes of the exciton are faster than its radiative decay, as can be seen from Figure 5a, the possibility of these non-radiative events is reduced in scaled-down QDs. Strain relaxation-related radial potential well confines the exciton away from the QD sidewall. Moreover, for a smaller-diameter QD, the sidewall surface area is drastically reduced, leading to a net reduction in surface states. Additionally, longer chemical and PEC etchings reduce surface damage as well as surface states[57-59]. As a consequence, the exciton majorly decays through the radiative channel. A single and fast exciton decay time constant of 260 ps was previously reported for an 8 nm-diameter $In_{0.14}Ga_{0.86}N$ QD in my previous work[36]. It also validates the above argument of having a single major radiative decay channel of the exciton. This is beneficial to achieve the smallest $g^{(2)}(0)$ reported in this work for sample S5, as shown in Figure 3b. Though this $g^{(2)}(0)$ is still high, which can be ascribed to background contamination from the defect peak, which could not be filtered spectrally. On the contrary, for samples S2 and S3 with 36 nm-diameter QDs, the Auger recombination rate is in the order of thermal and tunnelling rates of the biexciton, as evident from Figure 5a. Additionally, there is a finite probability of biexciton radiative decay. These cause spectral broadening of the biexciton peak, as evidenced by the larger spectral linewidths observed for samples S2 and S3, in Figure 3a. In these cases, the exciton radiative decay is also accompanied by its surface recombination events and, hence, spectral broadening of the exciton peaks. This is supported by experimentally observed higher exciton peak linewidths for S2 and S3 reported in Figure 3a. These factors lead to zero-delay multi-photon events for samples S2 and S3, as shown in Figure 3b. Inhomogeneous broadening, discussed so far, also reduces ρ, which signifies loss of single-photon purity as explained above in the context of Figure 3c.

### F. Calculation of $g^{(2)}(0)$ from Carrier Dynamics

Finally, $g^{(2)}(0)$ is calculated by feeding all relevant time constant values into equation (8). This diameter-dependent $g^{(2)}(0)$ calculated by analysing carrier dynamics is shown in Figure 5b. Directly measured $g^{(2)}(0)$ values and those calculated from μPL spectra of the samples are indicated by black rectangles and red circles, respectively, in the figure. The experimentally measured $g^{(2)}(0)$ is further corrected for background contamination. Details of the background correction process are provided in my previous work[36]. Background-corrected $g^{(2)}(0)$ results are indicated by blue rhombuses in the figure. Except for S5, the $g^{(2)}(0)$ values obtained from experiment and carrier dynamics calculation agree well, whereas the $g^{(2)}(0)$ values after a background correction closely match with those calculated from μPL data. For sample S5, a discrepancy between the experimental and theoretical $g^{(2)}(0)$ may be attributed to the instrument response function of the setup. The background correction scheme also did not give any better $g^{(2)}(0)$ in this case. It is worth noting that, except for S5, $g^{(2)}(0)$ values calculated using μPL data are less than those obtained from carrier dynamics analysis. This is expected, as ρ extracted from µPL data completely filters out background noise during $g^{(2)}(0)$ calculation, and hence it shows a good match with background-corrected $g^{(2)}(0)$. On the other hand, the background is taken into account in the carrier dynamics calculation through surface recombination terms. However, in the case of sample S5, the Auger term completely dominates the other terms, so the influence of surface recombination is almost negligible. This causes a $g^{(2)}(0)$ similar to that calculated from µPL. These results validate that SPE is possible only for QD diameters below 31 nm. Thus, for the current $In_{0.14}Ga_{0.86}N$ QD with 3 nm height, 31 nm can be considered the single-photon-limited diameter at room temperature.

### G. Calculation of $g_0^{(n)}$

The higher-order autocorrelation functions are calculated as a function of the background correction factor (ρ) at zero-delay to understand their importance in characterizing a multi-photon quantum emitter. After substituting $g^{(2)}(0)$ from equation (4), the higher-order functions can be written in terms of a second-order function as mentioned below[60],

$$g^{(3)}(0,0) = \frac{g^{(2)}(0)^2}{2-g^{(2)}(0)} = \frac{1-2\rho^2+\rho^4}{1+\rho^2} \tag{28}$$

$$g^{(4)}(0,0,0) = \frac{g^{(2)}(0)^3}{(3-2g^{(2)}(0))(2-g^{(2)}(0))} = \frac{(1-\rho^2)^3}{(1+2\rho^2)(1+\rho^2)} \tag{29}$$

These are plotted with ρ in Figure 5c, where experimentally measured QDs are marked by dashed lines. It is evident from the figure that $g_0^{(n)}$ reduces with an increase in order n for any particular ρ, where 0<ρ<1. For example, $g^{(2)}(0)=0.5$ corresponds to $g^{(3)}(0,0)= 0.167$ and $g^{(4)}(0,0,0)= 0.0417$. It is also important to note that $g^{(4)}(0,0,0)<0.5$ for a corresponding $g^{(2)}(0)\gg0.5$, thereby suggesting a finite multi-photon probability, as can be seen in the case of samples S2 and S3 in Figure 5c. The first-order derivative of $g_0^{(n)}$ is plotted with ρ in the inset. A vertical dashed line marks the single-photon limit of ρ. The right side of this line determines the SPE zone. It can be noted that the rate of change of $g_0^{(n)}$ increases with a decrease in n for any particular ρ and is maximum for the second-order in this zone. Hence, second-order autocorrelation is treated as the fundamental experiment to justify SPE from a quantum light source. Although second-order autocorrelation shows clear zero-delay antibunching for all QD samples in this study, it may exhibit bunching for n-photon Fock states for n>1 and is therefore misleading for quantitative analysis of such states of quantum light[61, 62]. The n$^{th}$ order autocorrelation employing n detectors in the setup is universal, as it can particularly characterize all such states for any n value.

## IV. CONCLUSIONS

In summary, laterally-etched $In_{0.14}Ga_{0.86}N$ QDs are fabricated by means of a combination of dry and wet etching techniques. QDs with a height of 3 nm and a diameter between 8 and 36 nm are reported. SPE is captured for QDs with diameters less than 31 nm. In contrast, shallow zero-delay antibunching with multi-photon emission probability is observed for QDs with a diameter of 36 nm. Thermal recombination of the exciton and unstable biexciton mediated by surface states is responsible for spectral broadening and hence the absence of any SPE in this range. Below 35 nm, the Auger recombination gradually becomes dominant with further reduction in QD diameter, thereby reducing spectral broadening and favouring SPE below 31 nm. DOS-corrected Auger recombination is reported to address the reduced DOS at the reduced dimension of QDs. A significant electron-hole spatial overlap below 9 nm QD diameter causes the formation of a stable biexciton. This biexciton recombines through a single Auger channel driven by its ultrafast decay rate and forms an exciton, which subsequently decays majorly through the radiative channel. This preferential radiative decay of the exciton is caused by two factors. Exciton is electrostatically shielded from QD sidewall surface states by means of a natural radial potential well. In addition, the number of surface states is much lower in chemically or photoelectrochemically treated and smaller-diameter QDs, thereby favouring a larger biexciton-state population. Suppression of the side-channels during the exciton and biexciton decay greatly improves the single-photon purity of the exciton peak. The $g^{(2)}(0)$ values of QDs are calculated separately from carrier dynamics and SNR of µPL spectra as a function of QD diameter. These results are, respectively, in good agreement with the experimentally measured and background-corrected $g^{(2)}(0)$. Diameter-dependent $g^{(2)}(0)$ clearly dictates the upper limit of QD diameter for SPE. Higher-order autocorrelations are discussed in the context of accurately characterizing multi-photon emission from a quantum light source. This work elucidates the physics underlying the design and development of a semiconductor QD for the deterministic generation of high-purity single photons at room temperature.

**FIGURES**

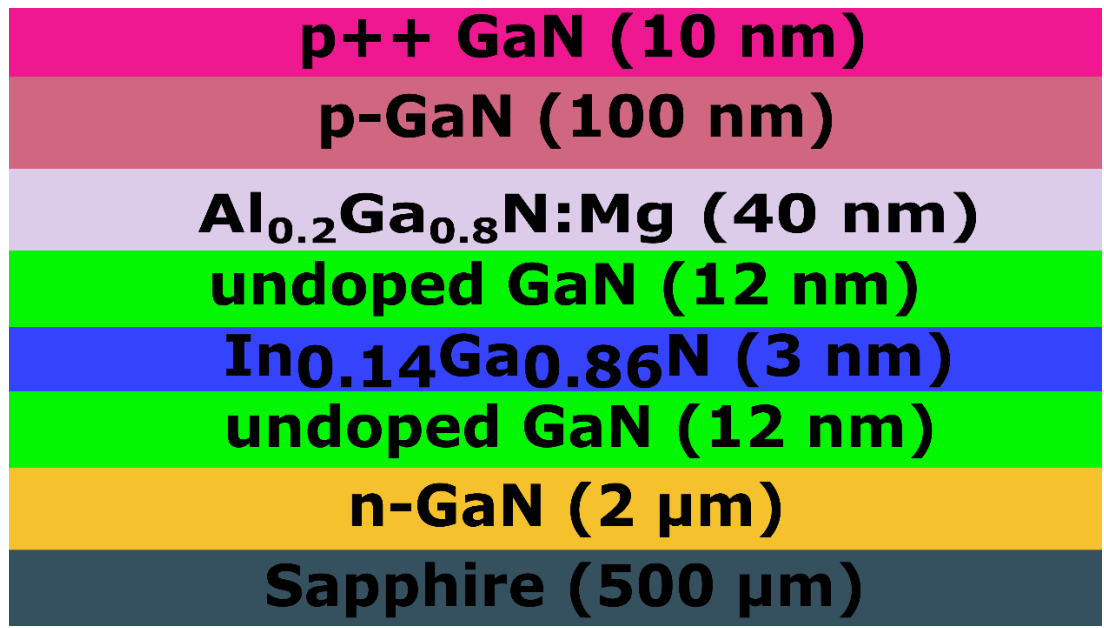

Figure 1. Schematic of InGaN/GaN heterostructure grown by MOCVD technique on a c-plane sapphire substrate.

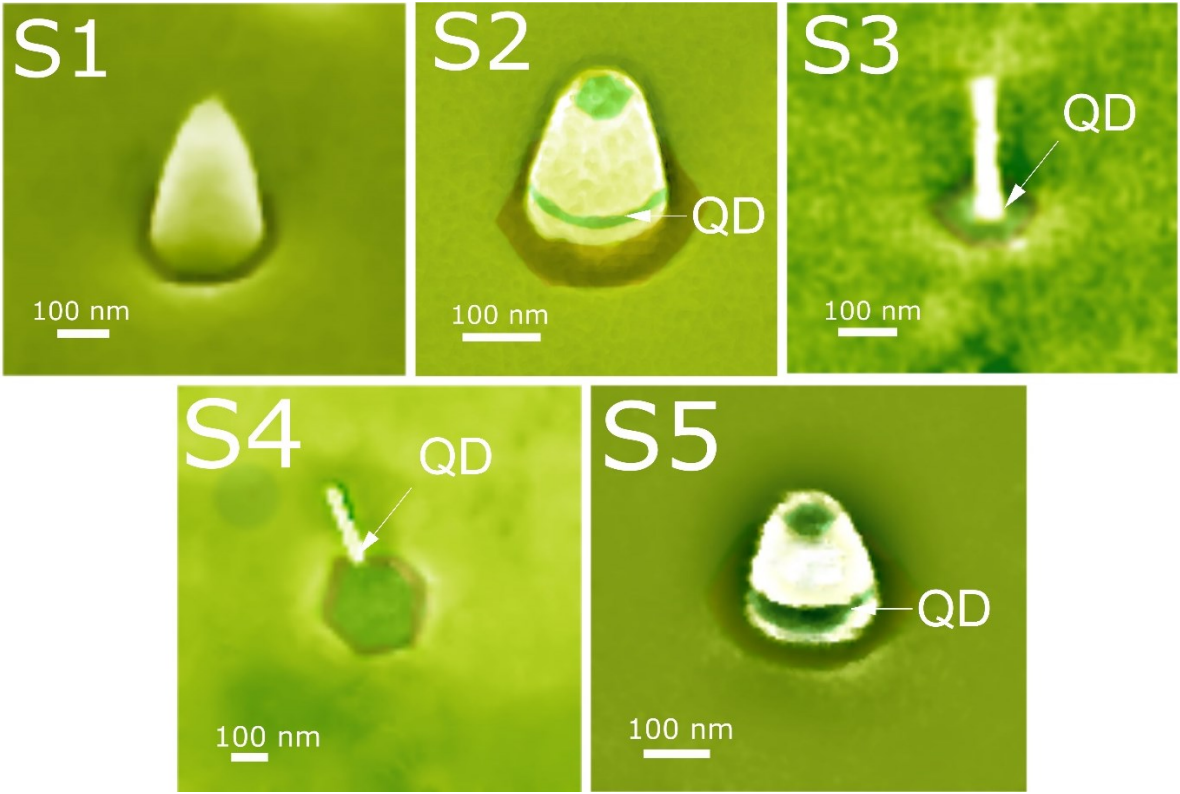

Figure 2. The cross-sectional SEM images of the samples. The location of QD in the nanowire is indicated by an arrow for each wet-etched sample

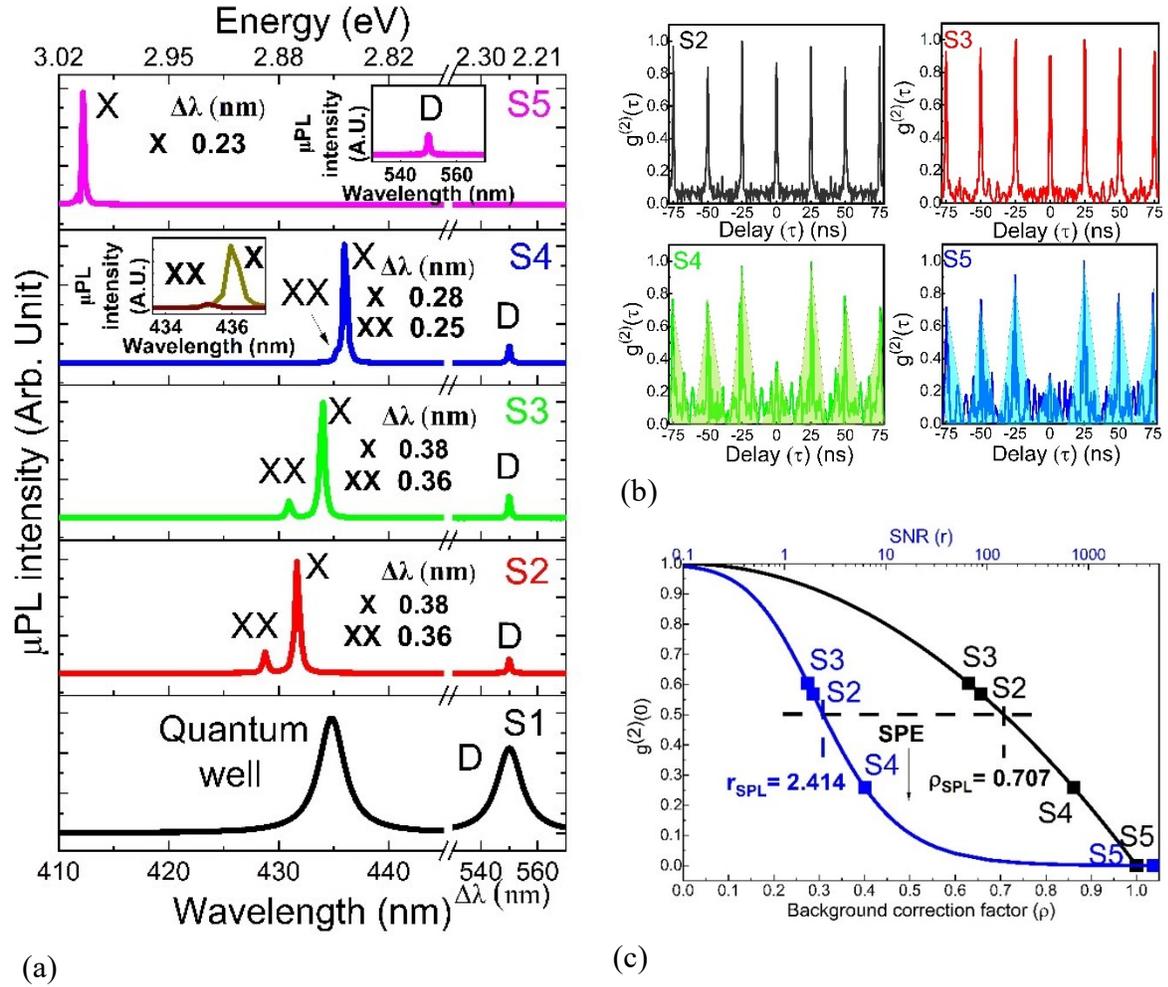

Figure 3. (a) µPL emission spectra of the samples show X, XX and defect-related peaks for S2, S3 and S4 QDs. Sample S5 shows X and defect-related (inset of S5) peaks. Lorentzian fits to the X and XX peaks for S4 are shown in the inset of panel S4. SNR increases, whereas linewidth decreases, with reduction in QD diameter, (b) room-temperature second-order autocorrelation for the exciton peak for samples S2-S5, $g^{(2)}(0)$ below 0.5 for samples S4 and S5 confirms SPE from these samples, (c) the plot of $g^{(2)}(0)$ as function of SNR and background correction factor, corresponding single-photon limits are marked by blue and black vertical dashed lines, $g^{(2)}(0)$ values calculated from µPL spectra are indicated by discrete points. SNR and background correction factor increase at lower dimensions, thereby improving single-photon purity.

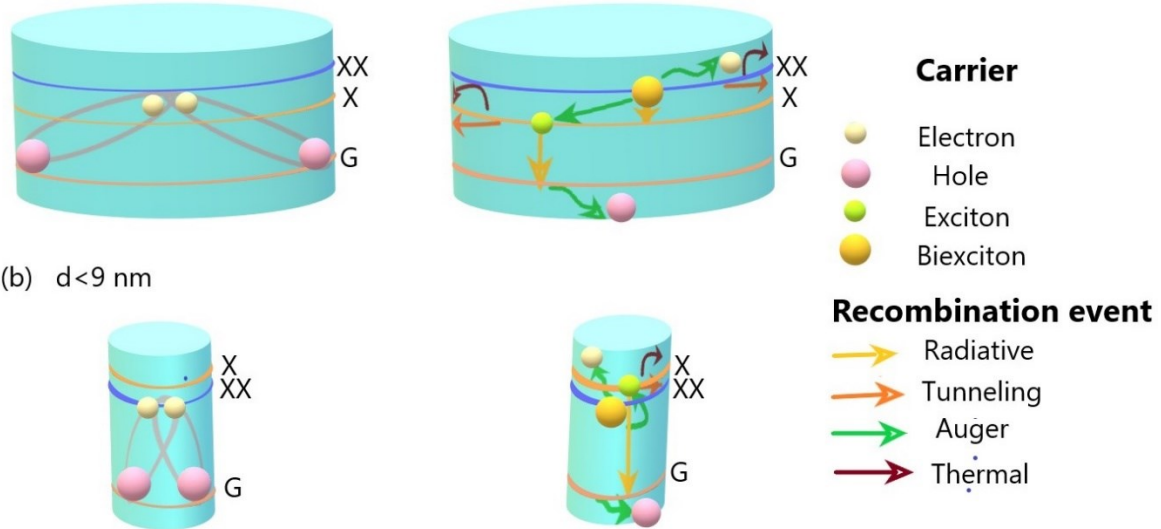

Figure 4. The schematic of radiative and principal non-radiative processes of exciton and biexciton in (a) larger (diameter>35 nm) and (b) smaller-diameter (diameter<9 nm) QDs. For smaller-diameter QDs, e-h spatial overlap is larger, and biexciton decays only via the Auger process to form an exciton, which subsequently decays to the ground state via radiative and non-radiative paths.

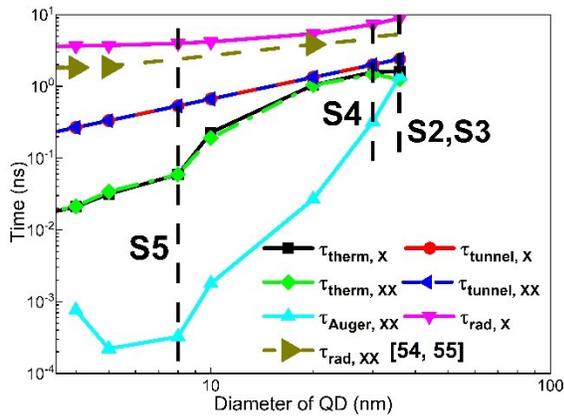
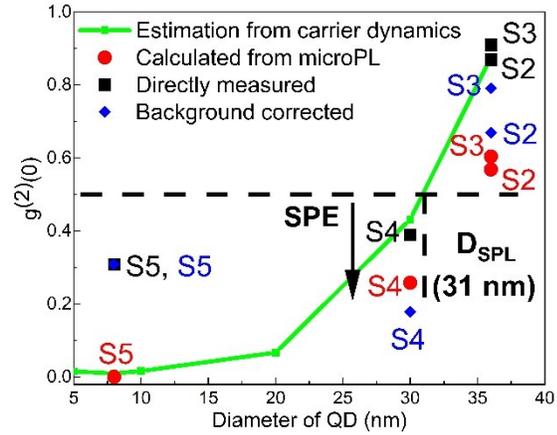

(a) (b)

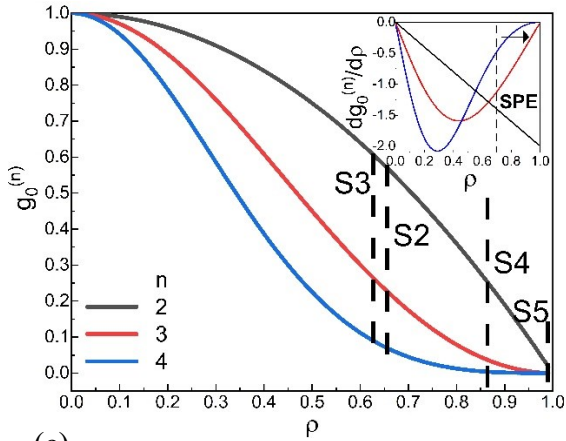

(c)

Figure 5. (a) diameter-dependent radiative and non-radiative lifetimes of exciton and biexciton are shown, diameters of experimentally measured QDs are indicated by dashed lines, Auger recombination is the dominant mechanism for samples S4 and S5, (b) $g^{(2)}(0)$ calculated using carrier dynamics is plotted as a function of QD diameter, experimentally measured $g^{(2)}(0)$, background-corrected $g^{(2)}(0)$ and $g^{(2)}(0)$ calculated using µPL data are shown by black squares, blue rhombuses and red circles, respectively, experimental $g^{(2)}(0)$ results agree well with those calculated from carrier dynamics, whereas background-corrected results closely match with $g^{(2)}(0)$ obtained from µPL analysis. The horizontal dashed line indicates the maximum limit for single photon emission, which suggests that samples S4 and S5 are single-photon emitters whereas, S2 and S3 are multi-photon emitters, single-photon limited diameter ($D_{SPL}$) for room-temperature SPE is marked by vertical

dashed line, single-photon emission is expected only left to this line, (c) higher-order autocorrelations at zero-delay are plotted with background correction factor ($\rho$), QDs used in the present work are indicated by dashed lines, first order derivatives of $g^{(2)}(0)$ are shown as function of $\rho$ in the inset where SPE occurs only on the right side of the dashed line, second-order autocorrelation shows the maximum sensitivity in single-photon domain making it the best choice as single-photon detection experiment.

## Acknowledgements

The author is grateful to the Indian Institute of Technology Bombay Nanofabrication facility and characterization labs for the major support. The author is extremely thankful to his PhD supervisor, Prof. Dipankar Saha, for providing all experimental support.

4. Ł. Dusanowski, S. H. Kwon, C. Schneider, and S. Höfling, "Near-Unity Indistinguishability Single Photon Source for Large-Scale Integrated Quantum Optics," Phys. Rev. Lett., 122 (17), 173602 (2019).

5. R. Uppu, H. T. Eriksen, H. Thyrrestrup, A. D. Uğurlu, Y. Wang, S. Scholz, A. D. Wieck, A. Ludwig, M. C. Löbl, R. J. Warburton, and P. Lodahl, "On-Chip Deterministic Operation of Quantum Dots in Dual-Mode Waveguides for a Plug-and-Play Single-Photon Source," Nat. Commun., 11 (1), 1-6 (2020).

6. S. Kolatschek, C. Nawrath, S. Bauer, J. Huang, J. Fischer, R. Sittig, M. Jetter, S. L. Portalupi, and P. Michler, "Bright Purcell Enhanced Single-Photon Source in the Telecom O-Band Based on a Quantum Dot in a Circular Bragg Grating," Nano Lett., 21 (18), 7740-7745 (2021).

7. P. Barigelli, F. Sirovich, G. Carvacho, and F. Sciarrino, "Rotational-Invariant Quantum Key Distribution Based on a Quantum Dot Source," Opt. Quantum., 3 (6), 518-524 (2025).

8. C. Y. Lu, and J. W. Pan, "Quantum-Dot Single-Photon Sources for the Quantum Internet," Nat. Nanotechnol., 16 (12), 1294-1296 (2021).

9. S. W. Xu, Y. M. Wei, R. B. Su, X. S. Li, P. N. Huang, S. F. Liu, X. Y. Huang, Y. Yu, J. Liu, and X. H. Wang, "Bright Single-Photon Sources in the Telecom Band by Deterministically Coupling Single Quantum Dots to a Hybrid Circular Bragg Resonator," Photonics Res., 10 (8), B1-B6 (2022).
22